\documentclass{elsart1p}
\usepackage{graphicx}
\usepackage{amssymb}
\begin{document}
\begin{frontmatter}
%
%
%
%
%
\title{Theory of Heavy Flavor in the Quark-Gluon Plasma}
%
%

\author{Ralf Rapp}

\address{Cyclotron Institute and Department of Physics and Astronomy, 
Texas A\&M University, College Station, TX 77843-3366, U.S.A.}

\begin{abstract}
Heavy-quark interactions in the Quark-Gluon Plasma are analyzed 
in terms of a selfconsistent Brueckner scheme using a thermodynamic 
$T$-matrix based on a potential model. The interrelations between
quarkonium correlators, spectral functions and zero-modes, and
open heavy-flavor transport and susceptibilities are elaborated.
Independent constraints from thermal lattice QCD can be used to
improve predictions for heavy-quark phenomenology in heavy-ion 
collisions.   
\end{abstract}

\end{frontmatter}

\section{Introduction}
\label{sec_intro}
\begin{figure}[!t]
\centering
\includegraphics[width=0.49\textwidth]{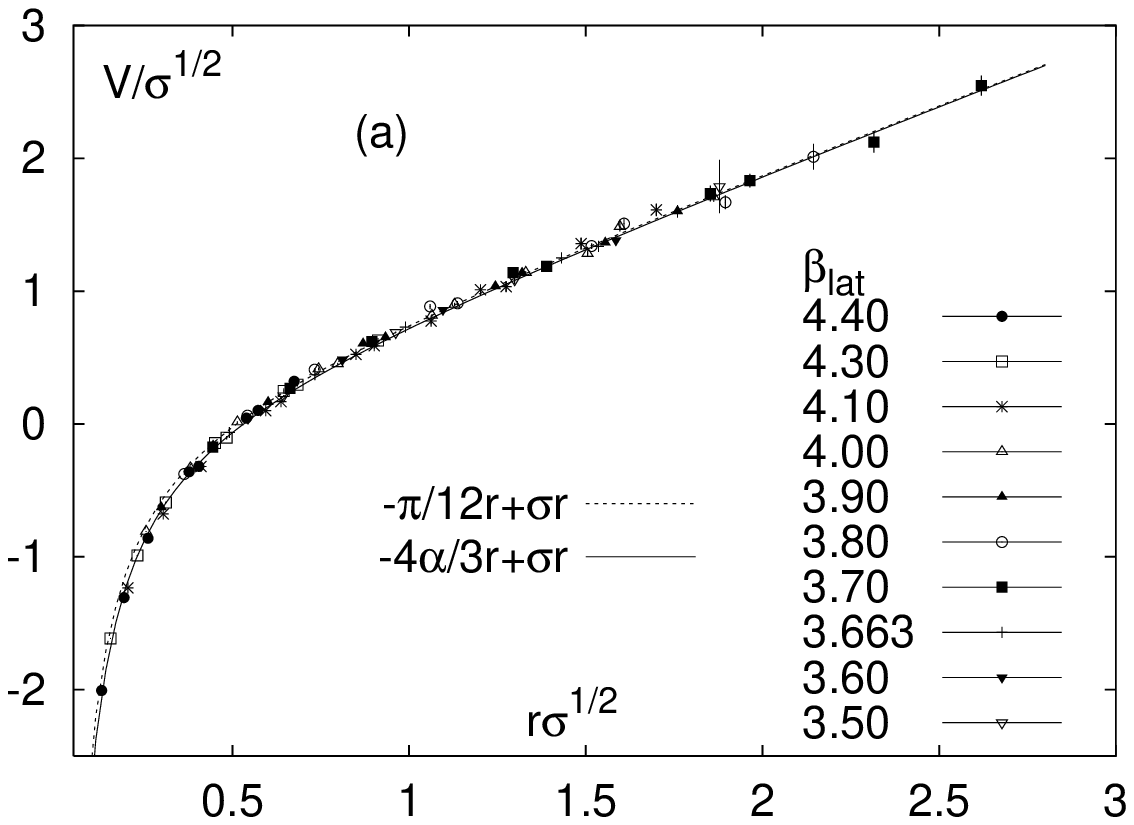}
\includegraphics[width=0.49\textwidth]{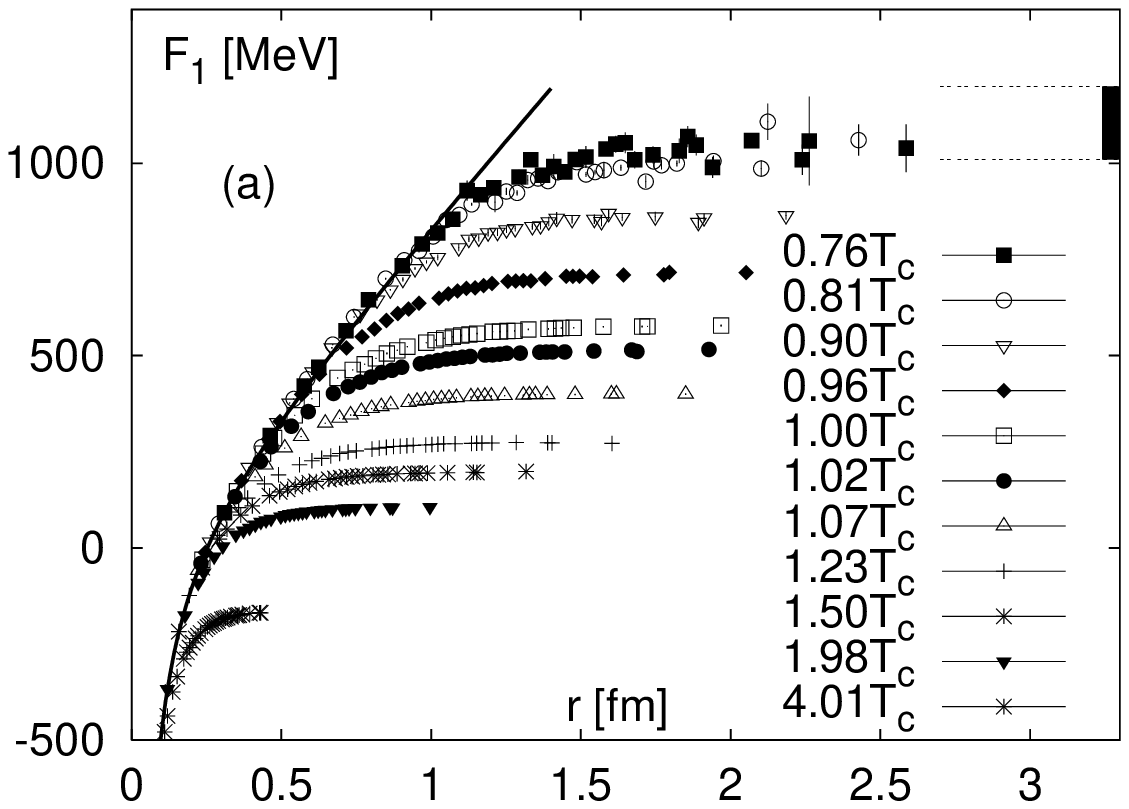}
\caption[]{The static heavy-quark potential in vacuum 
(left)\cite{Karsch:2000kv}
and the color-singlet free energy in medium (right; reprinted with
permission from \cite{Kaczmarek:2005ui}) as ``measured" in lattice QCD 
as a function of $Q$-$\bar Q$ separation.}
\label{fig_fqq}
\end{figure}

A basic challenge in many-body physics is the understanding of matter 
properties in terms of the forces between the constituents. Medium 
modifications of the force (or potential) render its determination an 
additional challenge. It is therefore important to have a good control 
over the force at least in the vacuum. In Quantum Chromodynamics (QCD), 
the fundamental force between static charges, i.e., a heavy quark ($Q$) 
and antiquark ($\bar Q$), is well established, both theoretically and 
phenomenologically. The heavy-quark (HQ) potential has been extracted 
with high accuracy from lattice-QCD (lQCD) computations~\cite{Karsch:2000kv} 
and can be well represented by the so-called Cornell potential, 
\begin{equation}
V_{Q\bar Q}(r) = -\frac{4}{3} \frac{\alpha_s}{r} + \sigma r \ , 
\end{equation} 
characterized by a perturbative color-Coulomb interaction at small 
distances and a ``string" term dominant at large $r$, cf.~left panel of 
Fig.~\ref{fig_fqq}.
This potential successfully describes charmonium and bottomonium 
spectroscopy in vacuum, which can be understood as an effective field 
theory (EFT) of QCD in a $1/m_Q$ expansion ($m_Q$: HQ mass). The string 
tension, $\sigma \simeq 1$\,GeV/fm, is of nonperturbative origin and most 
likely associated with the gluon-condensate structure of the QCD vacuum. 
The string term plays an important role already at rather small distances;
e.g., for $V(r_0)$=0, i.e., at $r_0$$\simeq$$\frac{1}{4}$\,fm, it is equal 
in magnitude (but opposite in sign) to the Coulomb term. Consequently, 
the charmonium spectrum is largely governed by the nonperturbative force 
(e.g., the ground-state binding, $E_B^{J/\psi}$$\simeq$\,0.6\,GeV, 
collapses to $\sim$0.05\,GeV if the string term is switched off).       
With a ``calibrated" strong force in vacuum at hand one can study its 
medium modifications and infer from these information about the 
phase structure and transport properties of QCD matter. The analysis
of quarkonium production and heavy-flavor spectra in ultrarelativistic 
heavy-ion collisions (URHICs) is aimed at precisely at these objectives, 
i.e., to identify signals of deconfinement and to extract heavy-quark 
diffusion coefficients from the produced medium (see, e.g., 
Refs.~\cite{Rapp:2008tf,BraunMunzinger:2009ih,Kluberg:2009wc,Rapp:2009my}
for recent reviews). 

The HQ free energy has also been computed at finite temperature, 
cf.~right panel of Fig.~\ref{fig_fqq}. One observes a gradual 
penetration of medium effects to smaller distances, naturally 
interpreted as a decrease of the ``Debye" screening length, 
$r_D \sim 1/m_D$ ($m_D$: Debye mass).
However, even at temperatures as high as 
$2T_c\approx$\,350\,MeV, the free energy still levels off at a positive 
value indicative for nonperturbative effects (string term).  
The applicability of the potential approach requires the 4-momentum 
transfer to be dominantly spacelike, 
i.e., $q_0\simeq {\vec q\,}^2/2m_Q \ll |\vec q|$. In the
vacuum this is satisfied by the smallness of the quarkonium binding 
energy. In the medium the latter is expected to decrease further. 
The thermal momentum scale of a single heavy quark also remains 
parametrically large not too far above $T_c$,   
$p_{\rm th}^2 \simeq  2m_QT \gg T^2$ ($T^2$: momentum transfer from 
the medium). Thus, thermal $Q$-$\bar Q$ and $Q$-medium interactions 
are essentially static and elastic but involve nonperturbative 
interactions. This suggests the possibility of a unified description
of heavy quarkonia and heavy-flavor transport, with a simultaneous  
treatment of bound and scattering states including resummations. The 
thermodynamic $T$-matrix approach, 
which is based on potential interactions, is such a framework, and has 
been successfully applied to electromagnetic plasmas~\cite{Redmer:1997}. 
The in-medium QCD interaction is, of course, much more involved than in
QED, but the idea of using input from lQCD has revived the potential 
approach in recent years~\cite{Mocsy:2007bk,Rapp:2008tf}. 
In this paper we will elaborate on the 
$T$-matrix approach for open and hidden heavy flavor in the QGP
(Sec.~\ref{sec_12body}), and how numerical results for spectral 
and correlation functions, as well as transport coefficients, can
be tested by thermal lQCD (Sec.~\ref{sec_tmat}).
We conclude in Sec.~\ref{sec_concl}. 
 
\section{One- and Two-Body Correlations in the QGP}
\label{sec_12body}
The commonly studied quantity in thermal lattice QCD characterizing
the propagation of a hadronic current with quantum numbers $\alpha$ 
is the imaginary-time ($\tau$) correlation function which is given
by a thermal expectation value as
\begin{equation}
G_\alpha(\tau,\vec r) = 
\langle\langle j_\alpha(\tau,\vec r) 
j^\dagger_\alpha(0,\vec 0)\rangle\rangle \  
= \  \parbox{1.2cm}{\includegraphics[scale=0.5]
{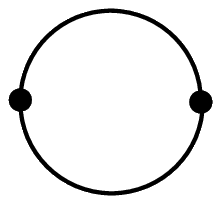}}+ \ \parbox{2.5cm}
{\includegraphics[scale=0.5]{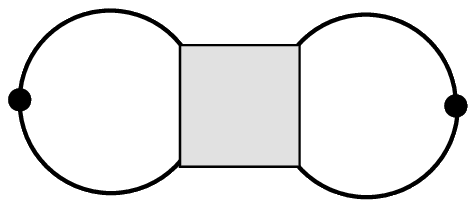}} \ . 
\label{Gtau}
\end{equation}   
The second equality is a diagrammatic representation for
a meson state in terms of its free quark-antiquark
loop and a 2-body interaction term.  
The physical information on the excitation spectrum in that channel
is given by the spectral function in momentum space, 
$\rho_\alpha(E,p) = -2~{\rm Im}~G_\alpha^R(E,p)$. 
It is related to the euclidean correlator, Eq.~(\ref{Gtau}), via
\begin{equation}
G_\alpha(\tau,p;T) = \int\limits_0^\infty \frac{dE}{2\pi} 
\rho_\alpha(E,p;T) \frac{\cosh[E(\tau-1/2T)]}{\sinh[E/2T]} \ ,  
\end{equation}
which illustrates the difficulty in extracting spectral functions
from lQCD ``data" of the euclidean correlator, since the latter is
only obtained for a finite number of $\tau$-points on a finite interval, 
0$<$$\tau$$<$$\frac{1}{2T}$. However, using model calculations for the 
spectral function a straightforward comparison to
lQCD data can be performed and constraints evaluated. Note that the
energy integration encompasses both bound and continuum states.
Within a HQ potential framework, the $T$-matrix formalism is thus 
an ideal choice~\cite{Mannarelli:2005pz,Cabrera:2006wh,Riek:2010fk}.   

Let us first focus on the quarkonium ($Q\bar Q$) sector. In ladder
approximation, and after partial-wave expansion, the in-medium 
$T$-matrix takes the form~\cite{Cabrera:2006wh,Riek:2010fk}
\begin{equation}
T_{\alpha}(E;q^\prime,q)=\mathcal{V}_{\alpha}(q^\prime,q) +\frac{2}{\pi}
\int\limits^\infty_0 dk\,k^2\,\mathcal{V}_{\alpha}(q^\prime,k) \,
G_{Q\bar Q}^0(E;k)\,T_{\alpha}(E;k,q) \ , 
\label{Tmat}
\end{equation}
where $\mathcal{V_\alpha}$ denotes the momentum-space potential and 
$G_{Q\bar Q}^0$ the uncorrelated 2-particle propagator (the first 
diagram in Eq.~(\ref{Gtau})); its imaginary part can be 
expressed via the in-medium 
single quark and anti-quark spectral functions, $\rho_{Q}$ and 
$\rho_{\bar Q}$, as~\cite{Riek:2010py}
\begin{eqnarray}
{\rm Im}~G_{Q\bar Q}^{0}(E,k)&=& - \int \frac{d\omega}{2\pi}
\left(\rho_Q(\omega,k) \rho_{\bar Q}(E-\omega,k) 
[1-f^Q(\omega)-f^{\bar Q}(E-\omega)]  \right. 
\nonumber\\
 & & \qquad \qquad  \left. +  \rho_Q(\omega,k) \rho_Q(E+\omega,k)
[ f^Q(\omega,k) - f^Q(E+\omega,k) ] \right) \ .
\label{ImG}
\end{eqnarray} 
The first term characterizes the standard uncorrelated $Q\bar Q$ 
propagation, while the second term (arising from negative-energy 
contributions) represents $Q\to Q$ (or $\bar{Q}\to \bar{Q}$) scattering 
which is nothing but the widely discussed zero-mode contribution to 
quarkonium correlators~\cite{Aarts:2005hg,Umeda:2007hy}. The scattering
effect is encoded in the medium modifications of the single-quark
spectral function, and thus intimately related to HQ transport.  
In the vector channel ($\alpha$=$V$), the $\mu\nu=00$ component 
represents the density-density correlation ($j_0$: density)
which in the static limit gives the HQ-number susceptibility,
\begin{equation}
\chi_{c}(T)=-\frac{\partial^2 \Omega}{\partial \mu_c^2} 
= \frac{\partial n_c}{\partial \mu_c} 
= \frac{1}{T}\int\limits_0^\infty \frac{dE}{2\pi}
\frac{2}{1-\exp(-E/T)}\rho_V^{00}(E)\ ,  
\label{chic}
\end{equation}
which is also governed by the zero mode. 

\begin{figure}[!t]
\centering
\includegraphics[width=0.75\textwidth]{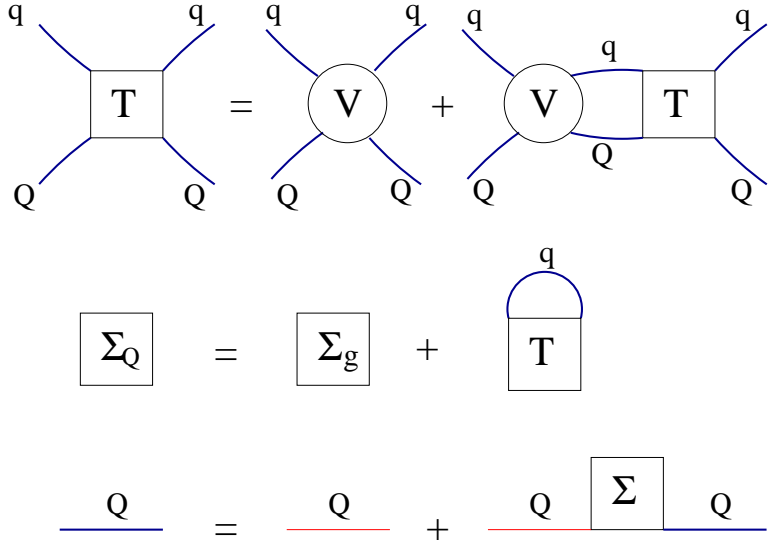}
\caption[]{Selfconsistent Brueckner problem for heavy quarks in the QGP:
the heavy-light quark $T$-matrix (upper panel) figures into the
calculation of the HQ selfenergy (middle panel) and propagator (lower
panel), which in turn enters into the 2-particle propagator of
the $T$matrix.}
\label{fig_brueck}
\end{figure}
The HQ spectral function is determined by the HQ selfenergy in the QGP, 
$\Sigma_Q$, as
\begin{equation}
\rho_Q(\omega,k)=\frac{-1}{\omega_Q(k)} 
{\rm Im}\frac{1}{\omega-\omega_{Q}(k)-\Sigma_{Q}(\omega,k)} \ , 
\end{equation}
cf.~the lower panel of Fig.~\ref{fig_brueck}. The selfenergy receives
contributions from the interactions of the heavy quark with the 
heat-bath particles, $\Sigma_{Q} \sim \int f^q T_{qQ}$,  but also from 
possible condensate remnants, especially gluon condensates (recall 
that the string tension, $\sigma$, seems to survive above $T_c$), as 
illustrated in the middle panel of Fig.~\ref{fig_brueck}. Following the 
arguments given in the Introduction, the $T$-matrix approach may also 
apply to heavy-light interactions. In this case one can establish a 
selfconsistent Brueckner scheme in which quarkonium and open heavy-flavor  
interactions follow from the same potential (the construction of such a 
potential will be discussed below).  
Heavy-quark rescattering in the QGP is, of course, at the origin of its
transport properties. The large HQ mass implies that momentum transfers 
are small compared to its thermal momenta, $q\ll p_{\rm th}$. This leads
to a Brownian motion picture with a Fokker-Planck equation for the 
time evolution of the HQ distribution function, schematically given by
\begin{equation}
  \frac{\partial f}{\partial t}=\gamma \frac{\partial (pf)}{\partial p}
  + D \frac{\partial^2f}{\partial p^2} \quad , \quad  
\gamma \, p \sim \int f^q \, |T_{Qq}|^2 \, (1-\cos\theta) \ . 
\label{fp}
\end{equation}
The thermal relaxation rate, $\gamma$, is closely related to the 
scattering rate $\Gamma_Q =-2\,{\rm Im}\,\Sigma_Q$; while the latter is 
computed with the forward scattering amplitude, the former includes a 
weight with the scattering angle, $\theta$, signifying the effect of 
isotropization. 
   
Let us briefly discuss the construction of the interaction potential,
$\mathcal{V}$, in Eq.~(\ref{Tmat}). Originally it was hoped to extract 
it model-independently from lQCD measurements of the HQ free
energy at finite $T$ (right panel of Fig.~\ref{fig_fqq}), in 
analogy to the vacuum case (left panel of Fig.~\ref{fig_fqq}). However, 
at finite $T$, the entropy contribution in $F = U - TS$ causes an 
ambiguity as to whether the free ($F$) or internal energy ($U$) should
be used. This problem is related to the interplay of time scales for 
the HQ interaction and for the thermal relaxation of the medium,
with $U$ and $F$ as limiting cases. On the other hand, we note 
that both $F$ and $U$ are computed from thermal expectation values, 
as differences between a system with and without an embedded  
$Q\bar{Q}$ pair. A more rigorous approach should
therefore start from (an ansatz for) a ``bare" potential (figuring into 
the $T$-matrix equation) and then calculate the free (and internal)
energies using suitable many-body techniques. The resulting
``zero-point" functions ($F$, $U$) can then be compared to lQCD data 
and the input 4-point function (potential) tuned for optimal agreement. 
The thus obtained potential can be employed to calculate 
heavy-quark and -quarkonium properties in the QGP as described above.     
    
\begin{figure}[!t]
\centering
\includegraphics[width=0.48\textwidth]{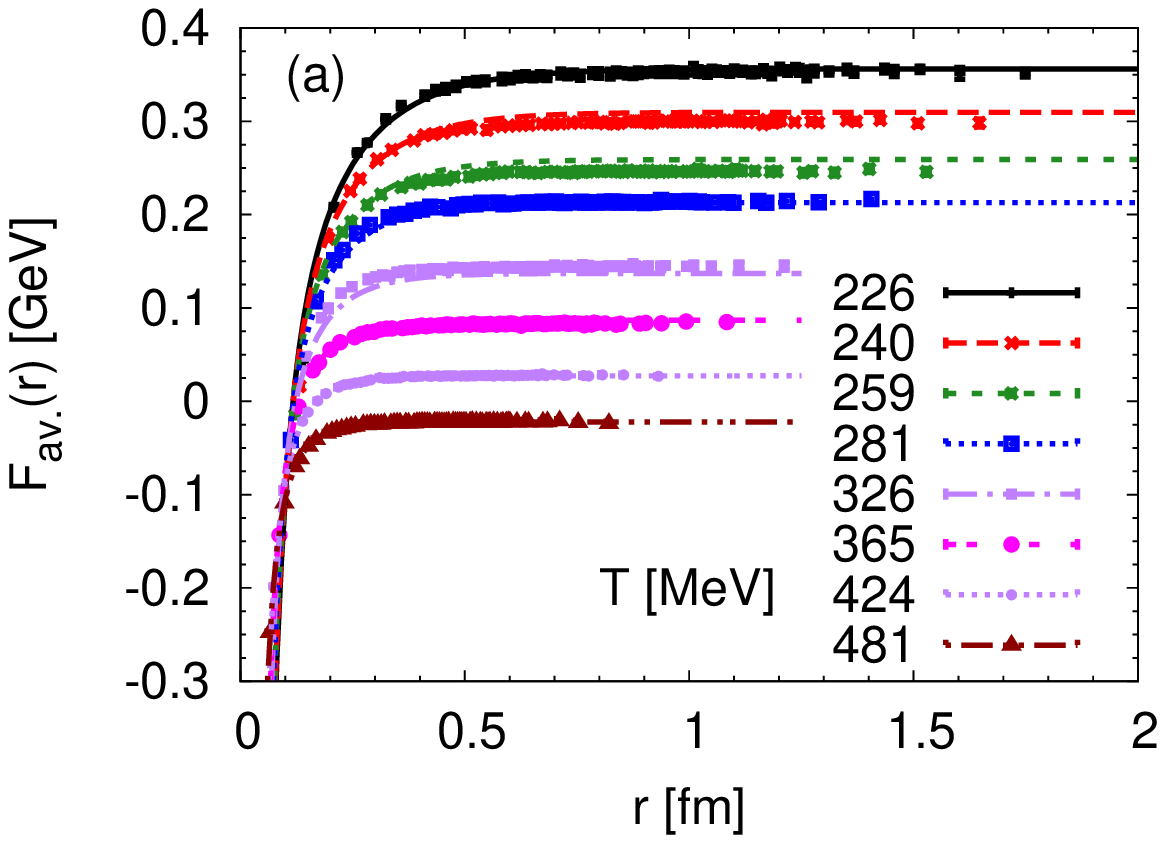}
\includegraphics[width=0.48\textwidth,height=0.335\textwidth]{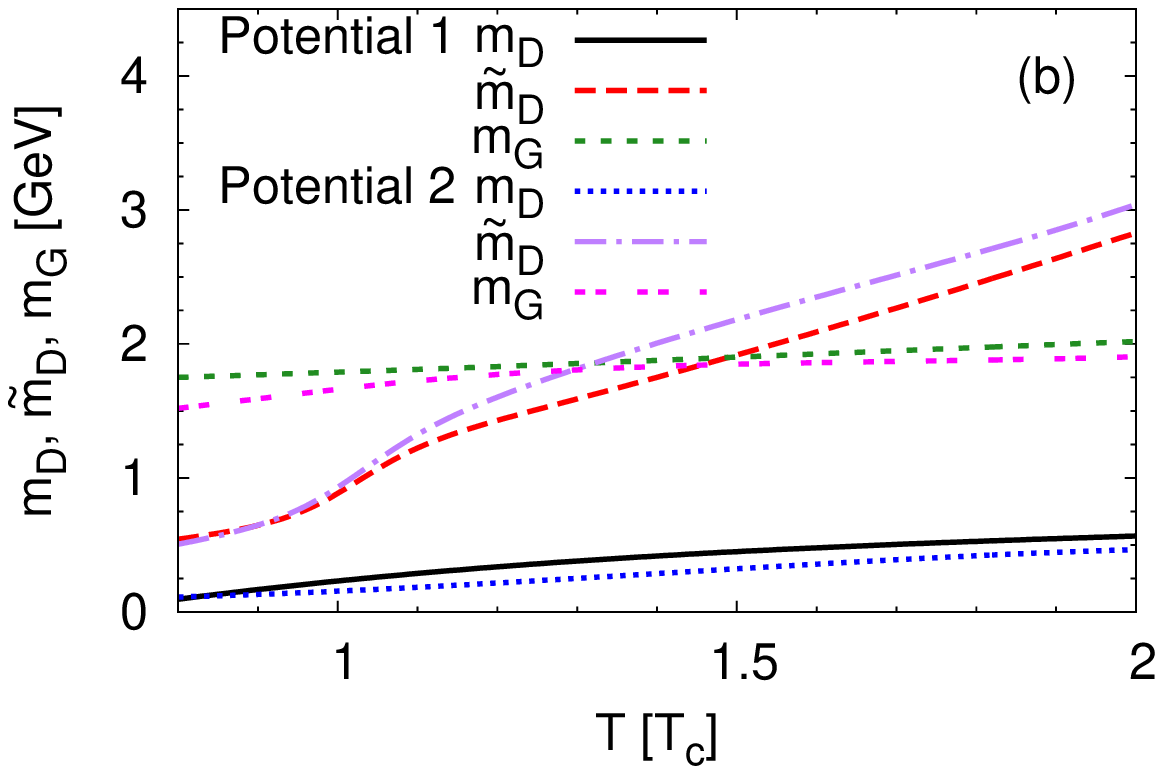}
\caption[]{Color-average $Q\bar Q$ free energy at finite $T$ (left panel) 
as computed in $N_f=2+1$ lQCD (symbols) and pertinent fits~\cite{Riek:2010fk}
using the ``Coulomb+string" ansatz, Eq.~(\ref{D00}). The $T$-dependence
of the fit parameters is displayed in the right panel for 2 different
lQCD inputs~\cite{Kaczmarek:2007pb,Petreczky:2004pz}. In both cases,
the $T$-dependence of $\alpha_s\simeq0.28-0.32$ turns out to be weak.
(Figures reproduced with permission from \cite{Riek:2010fk}.)}
\label{fig_fqq-fit}
\end{figure}
The free and internal energies are believed to represent limiting
cases bracketing the uncertainty in potential models.
Instead of a functional parameterization of the lattice results a 
field-theoretic approach accounting for the two main components of
the potential~\cite{Megias:2005ve} has been proven very 
useful. With an ansatz for the effective Coulomb + confining 
propagators (including couplings), 
\begin{equation}
D_{00}(k;T) = \frac{\alpha_s^2}{k^2+m_D^2} + 
              \frac{m_G^2}{(k^2+\tilde{m}_D^2)^2} \ ,  
\label{D00}
\end{equation}
four parameters characterizing the respective interaction strength 
($\alpha_s$, $m_G$) and in-medium screening ($m_D$, $\tilde{m}_D$)
can be adjusted to reproduce the finite-$T$ color-average free 
energy from lQCD fairly well (see Fig.~\ref{fig_fqq-fit}). To improve 
the applicability in the scattering regime a Breit correction as known 
from electrodynamics~\cite{Brown:1952} has been introduced, 
$V_{\rm Coul} \rightarrow V_{\rm Coul} (1-\vec{v}_1 \cdot\vec{v}_2)$,  
accounting for the magnetic current-current interaction. This
modification renders the Coulomb potential ``minimally" Poincar{\'e} 
invariant~\cite{Brambilla:2003nt} and the high-energy limit of the 
$T$-matrix consistent with the tree-level
perturbative QCD (pQCD) result~\cite{Riek:2010fk}. Note that this
extension only applies to a Lorentz-vector interaction (Coulomb 
term), not to the scalar confining term. It thus requires the 
explicit decomposition as given by the field-theoretic ansatz,
Eq.~(\ref{D00}). Further systematic improvements, e.g., scrutinizing
the uncertainty of retardation effects, need yet to be performed. 
Comparisons to EFT approaches at finite $T$~\cite{Brambilla:2008cx} 
could also prove useful, even though they are usually based on 
perturbative scale hierarchies which are problematic for the 
confining term. Interesting results are also emerging from lattice 
simulations of classical Yang-Mills theory~\cite{Laine:2009dd}, 
supporting the importance of nonperturbative effects.

\section{Brueckner Theory of Heavy Flavor in QGP}
\label{sec_tmat}
\begin{figure}[!t]
\centering
\includegraphics[width=0.5\textwidth,angle=-90]{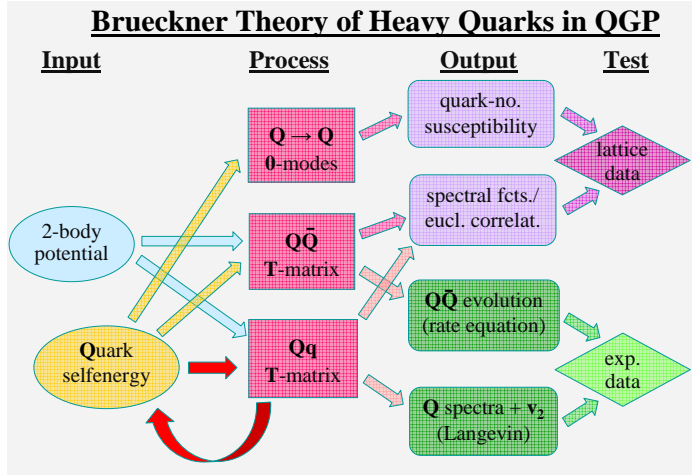}
\caption[]{Flow chart for a many-body approach to describe properties
to heavy quarks and quarkonia in the QGP with applications to lQCD
and heavy-ion data.}
\label{fig_chart}
\end{figure}
In Fig.~\ref{fig_chart} we display a flow chart envisioning a possible 
implementation of Brueckner theory for heavy quarks in the QGP. Starting 
from an ansatz for the bare $Q$-$\bar Q$ potential (with suitable 
corrections and extensions to the heavy-light sector), two-body 
scattering amplitudes are readily calculated. From the heavy-light
$T$-matrix one obtains a HQ selfenergy, which has to be
iterated for selfconsistency for use in the final $T$-matrices. On the 
one hand, the latter are employed to compute spectral and
correlation functions, susceptibilities and free/internal energies (not 
shown), all of which can be constrained by lattice ``data" (and  
used to narrow down the input potential). On the other hand, one 
calculates HQ transport coefficients and quarkonium properties (masses, 
lifetimes, dissociation temperatures) which are readily implemented 
into phenomenological descriptions of URHICs (e.g., Langevin simulations 
or rate equations in a hydrodynamically evolving bulk medium) and 
checked against 
experiment~\cite{vanHees:2007me,Gossiaux:2008jv,Alberico:2010tb,Das:2011fe,Zhao:2010nk,Liu:2010ej}.    

\begin{figure}[!t]
\centering
\includegraphics[width=0.485\textwidth]{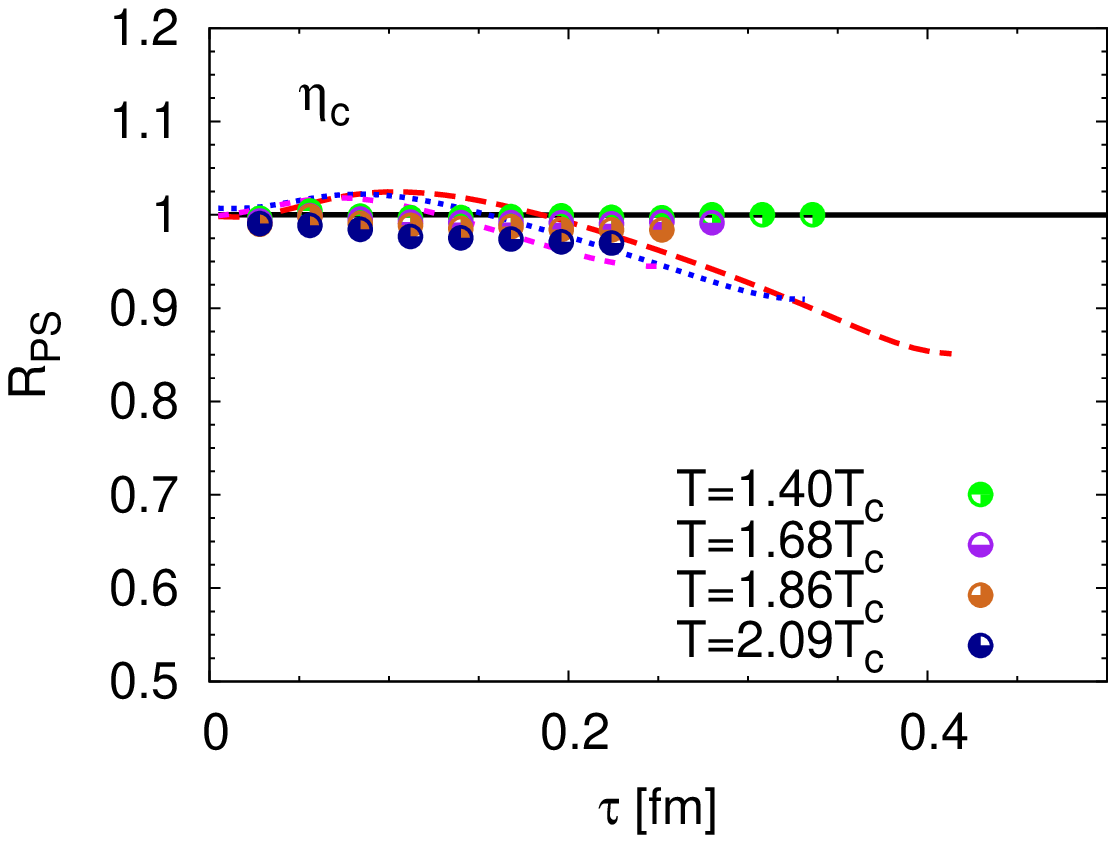}
\includegraphics[width=0.505\textwidth]{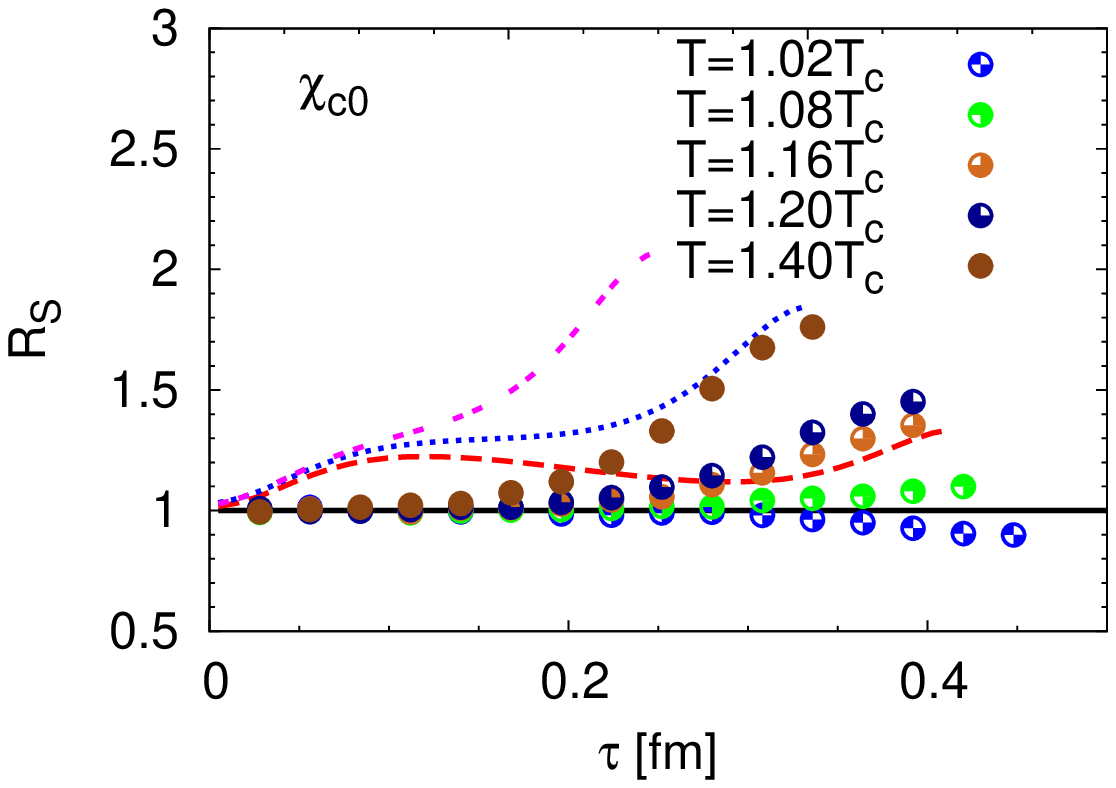}
\caption[]{Euclidean charmonium correlators ($G_\alpha$), normalized to 
the ones with vacuum spectal function but with finite-$T$ kernel,
$R_\alpha(\tau)\equiv G_\alpha(\tau)/G_\alpha^{\rm vac*}(\tau)$. The 
$T$-matrix results at $T$=1.2, 1.5 and 2\,$T_c$ (dashed, dotted and 
double-dashed curves)~\cite{Riek:2010py} are compared to 2-flavor lQCD 
computations~\cite{Aarts:2007pk} in the pseudoscalar ($\eta_c$) channel 
(left; no zero-mode) and scalar ($\chi_{c0}$) 
channel (right; including zero modes). }
\label{fig_charmonium}
\end{figure}
\begin{figure}[!t]
\centering
\includegraphics[width=0.5\textwidth]{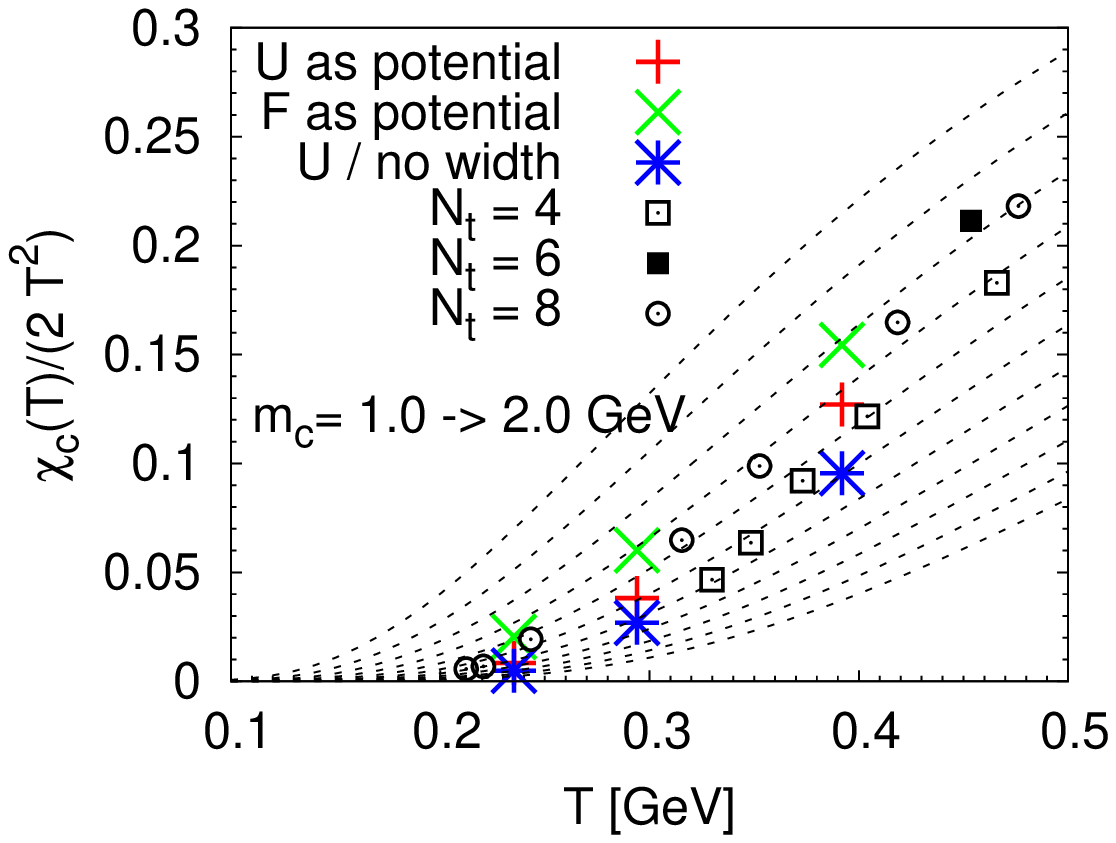}
\hspace{-0.4cm}
\includegraphics[width=0.51\textwidth]{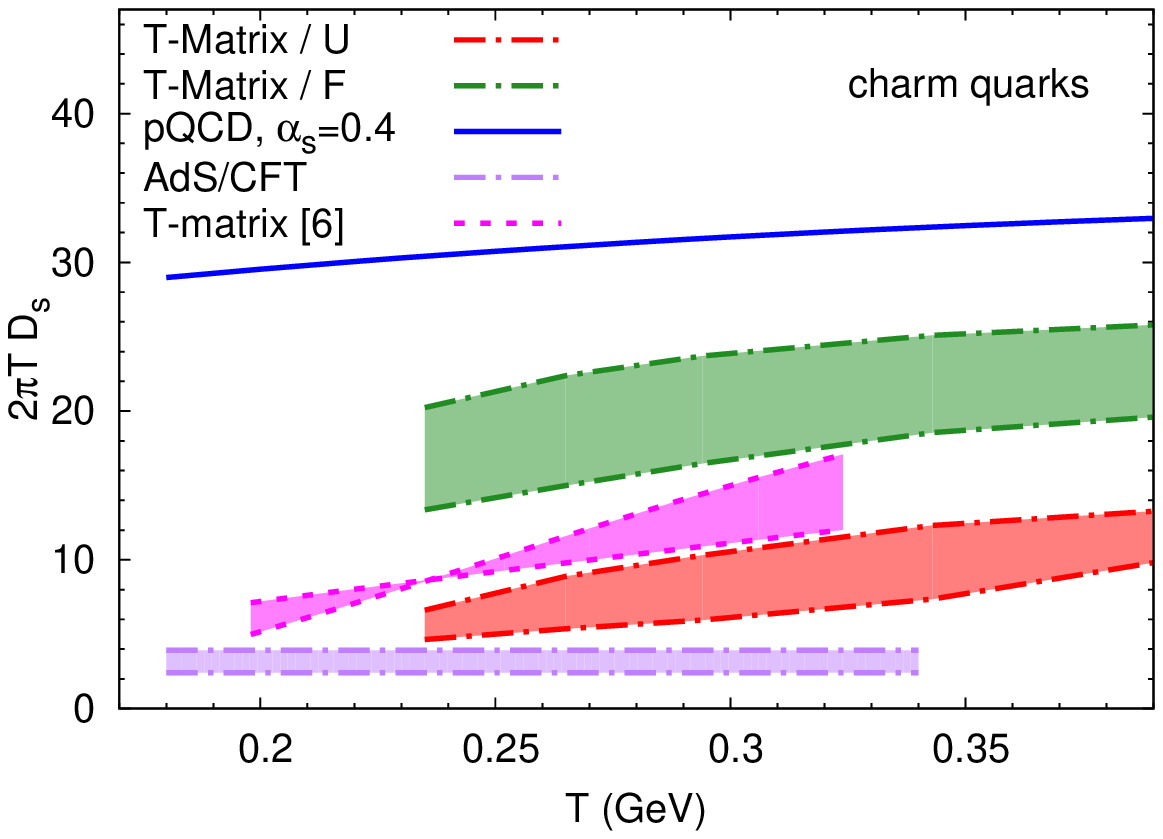}
\caption[]{Left: charm-quark susceptibility (normalized to twice the 
free massless limit) from selfconsistent Brueckner calculations 
with $U$ or $F$ potentials (colored symbols), lQCD 
(squares, circles)~\cite{Petreczky:2009cr} and non-interacting 
$c$-quarks with masses $m_c$=1-2\,GeV in steps of 0.1\,GeV (dashed 
lines, top down).
Right: spatial diffusion coefficient for $c$ quarks from 
the $T$-matrix with $U$ (pink~\cite{vanHees:2007me} 
and red bands~\cite{Riek:2010fk}) and $F$ potentials (green 
band)~\cite{Riek:2010fk}, LO pQCD (solid line) and
AdS/CFT matched to QCD~\cite{Gubser:2006qh} (lower band).}
\label{fig_hq-trans}
\end{figure}

As an example we show in Figs.~\ref{fig_charmonium} and \ref{fig_hq-trans} 
results from a recent selfconsistent Brueckner calculation including 
off-shell width effects in the HQ propagators~\cite{Riek:2010py}. When 
using the lQCD internal energy as input potential, one finds HQ widths 
of considerable magnitude, reaching up to $\Gamma_Q$$\simeq$0.1-0.2\,GeV 
for on-shell charm quarks. When implemented into the charmonium $T$-matrix,
the ground-state ($\eta_c$, $J/\psi$) melting temperature (estimated 
from the disappearance of the bound-state peak) is found to be around
1.5\,$T_c$, significantly smaller than in calculations neglecting width 
effects ($\sim$\,2\,$T_c$). The resulting euclidean correlators,
including zero modes according to Eq.~(\ref{ImG}), show fair agreement 
with lQCD data (see Fig.~\ref{fig_charmonium}), even though quantitative 
improvements are certainly warranted. Still within the same framework, 
the HQ susceptibility, Eq.~(\ref{chic}), has been computed and compared 
to lQCD data (left panel of Fig.~\ref{fig_hq-trans}). For this quantity
the width effects lead to an appreciable increase over the 
zero-width limit, providing better overlap with lQCD data. The results 
of a calculation with the free energy as potential generate a larger 
$\chi_c$, since for $F_{Q\bar Q}$ the in-medium HQ mass correction, 
$\Delta m_Q = F_{Q\bar Q}(r$$\to$$\infty)$, is about 0.3\,GeV 
smaller than for $U_{Q\bar Q}$. 
Finally, the resulting spatial HQ diffusion coefficients, computed from 
the selfconsistent heavy-light $T$-matrix within a Fokker-Planck 
equation (\ref{fp}) with $D_s = T/(m_Q\gamma)$, are displayed in 
Fig.~\ref{fig_hq-trans}. When using the $U$-potential, $D_s$ turns out 
to be a factor of 3-5 smaller (indicating stronger coupling) compared to 
pQCD. Toward $T_c$, the values are not very far from the strong-coupling 
limit represented by AdS/CFT (close to $T_c$ the $T$-matrix is 
not yet reliable, e.g., due to the lack of coupled channels such as 
$D\bar D$). With $F$ as potential $D_s$ is roughly in 
between the $U$-scenario and pQCD.

\section{Conclusions}
\label{sec_concl}
We have argued that nonperturbative elastic interactions
are key to understanding low-momentum interactions of heavy quarks
in the QGP. A potential-based $T$-matrix approach is, in principle,
capable of connecting heavy quarkonium and heavy-flavor transport
physics. This opens a rich arsenal of constraints from thermal
lattice QCD on quantities which directly relate to phenomenological
applications in URHICs. Many open question remain, e.g., a proper 
potential definition or the role of correlations near $T_c$ and 
nonperturbative $Q$-gluon interactions. Systematically addressing these 
issies will be essential for a full exploitation of upcoming 
high-precision heavy-flavor measurements at RHIC, LHC and FAIR.  

\vspace{0.3cm}

\noindent
{\bf Acknowledgment}
I thank the organizers of ICPAQGP10 for a very stimulating meeting.
I am indebted to my collaborators on the presented topics, D.~Cabrera,
V.~Greco, H.~van Hees, F.~Riek and X.~Zhao.
This work has been supported by the U.S. National Science Foundation
under grant no.~PHY-0969394 and by the A.v.~Humboldt Foundation (Germany).

\end{document}